\documentclass[letterpaper, 10 pt, conference]{ieeeconf}  

\IEEEoverridecommandlockouts                              

\usepackage{color}
\usepackage{multicol}
\usepackage[bookmarks=true]{hyperref}

\usepackage{gensymb}
\usepackage[pdftex]{graphicx}
\usepackage[cmex10]{amsmath}
\usepackage[ruled,vlined]{algorithm2e}
\usepackage{array}
\usepackage[caption=false,font=footnotesize]{subfig}
\usepackage{url}

\usepackage{amsthm}
\usepackage{amssymb}
\usepackage{listings}
\usepackage{tikz}
\usetikzlibrary{automata,positioning,shapes,arrows}
\usepackage{stmaryrd}
\usepackage{mathtools}

\graphicspath{{./pics/}}
\DeclareGraphicsExtensions{.pdf,.jpeg,.png}

\newtheorem{thm}{Theorem}

\theoremstyle{definition}
\newtheorem{defn}{Definition}

\theoremstyle{exampstyle} \newtheorem{example}{Example}
\theoremstyle{exampstyle} \newtheorem{problem}{Problem}
\newenvironment{model}[1][htb]
{
	\begin{algorithm}[#1]%
}{\end{algorithm}}
\newenvironment{model*}[1][htb]
{
	\begin{algorithm*}[#1]%
	}{\end{algorithm*}}

\title{\LARGE \bf Safety Certified Cooperative Adaptive Cruise Control under Unreliable Inter-vehicle Communications*
}

\author{Rafael Rodrigues da Silva$^{1,2}$ and Hai Lin$^{1}$
\thanks{*The financial supports from NSF-CNS-1239222, NSF-CNS-1446288  and NSF- EECS-1253488 for this work are greatly acknowledged.}
	\thanks{$^{1}$ All authors are with Department of Electrical Engineering, University of Notre Dame, Notre Dame, IN 46556, USA.
		{\tt\small (rrodri17@nd.edu;~hlin1@nd.edu)}}
\thanks{$^{2}$ The first author would like to appreciate the scholarship support by CAPES/BR, BEX 13242/13-0}
}

\begin{document}

\maketitle
\thispagestyle{empty}
\pagestyle{empty}

\begin{abstract}
	Cooperative adaptive cruise control(CACC) system provides a great promise to significantly reduce traffic congestion while maintaining a high level of safety. Recent years have seen an increase of using formal methods in the analysis and design of cooperative adaptive cruise control systems. However, most existing results using formal methods usually assumed an ideal inter-vehicle communication, which is far from the real world situation. Hence, we are motivated to close the gap by explicitly considering non-deterministic time delay and packet dropout due to unreliable inter-vehicle communications. In particular, we consider a passive safety property, which requests a vehicle to avoid any collisions that can be considered as its fault. Under the assumption that the communication delay is bounded and we know the upper bound, we then formally verify the passivity safety of a class of hybrid CACC systems. This result allows us to define a safe control envelope that will guide the synthesis of control signals. Vehicles under the CACC within the safe control envelope are guaranteed to avoid active collisions. 
\end{abstract}

\section{Introduction}

The road transportation system is crucial to the economic because it provides a reliable passenger and goods traffic that moves the domestic and international trade. Hence, the traffic safety and capacity is of major concern. In the US, more than 35,000 people died in motor vehicle accidents in 2015 \cite{trafficsafety2015}. Moreover, the growing traffic demand in the US is faster than its expansion and causes an increasing congestion cost that reached 160 billions of dollars in 2014 \cite{schrank2015urban}. Those facts are motivating transportation agencies to consider non-traditional solutions such as the Intelligent Transportation System (ITS) to maximize the traffic throughput of existing transportation system as reducing the risk of accidents \cite{dey2016review}.

In particular, Adaptive Cruise Control (ACC) was introduced as a possible solution to those issues. ACC allows a shorter safe distance to the preceding vehicle by measuring the vehicle's relative distance and velocity and adjusting the speed accordingly. Although radar sensors are adequate to measure relative distance, to use the wheel speed measurement shared through the network can reduce the uncertainty inherent in radar measurements and further reduce the safe distance between vehicles. Thus, a lot of research activities in ITS have been devoted to the study of Cooperative Adaptive Cruise Control (CACC) \cite{ploeg2011design,loos2011adaptive,mitsch2012towards,loos2013efficiency,van2015safety,santini2015consensus,dey2016review} and the references therein.

CACC systems are ensured through two different approaches: the string stability analysis and hybrid system verification. The string stability analysis approaches consider the problem of finding a regulator that the vehicle acceleration, velocity or distance error will not be amplified for a given headway (i.e. relative distance divided by the vehicle speed). For example, in  \cite{ploeg2011design}, the authors proposed a regulator that considers communication delay and presented experimental results with a platooning of similarly adapted vehicles. A regulator with an online uncertainties estimation was proposed in \cite{van2015safety}. A consensus-based approach for a string stable regulator was proposed in \cite{santini2015consensus} to address both communication delays and package losses. 

The hybrid system verification approaches, on the other hand, model the CACC system as a hybrid system and investigate the problem of finding the safe distance between vehicles. Most existing work along this line assumed an ideal inter-vehicle communication network, see e.g., \cite{loos2011adaptive} and \cite{mitsch2012towards}. The package loss was considered in \cite{loos2013efficiency}, but assumed  that it can find a timeout based on a network model to ensure the safety by asking for the human assistance. Therefore, a verification approach that considers both communication delay and package loss without relying on human interventions has not been considered yet.

Hence, we are motivated to close the gap by explicitly considering non-deterministic time delay and packet dropout due to unreliable inter-vehicle communications. In particular, we aim to verify a passive safety of a control envelope for CACC systems in differential dynamic logic (d$\mathcal{L}$) for a highway scenario where each vehicle is equipped with relative distance and velocity sensors and inter-vehicle communication devices. The passive safety specification requires that the vehicle should avoid any collision with other vehicles in which can be considered its fault \cite{macek2009towards}. A control envelope is a class of control systems which d$\mathcal{L}$ could model and verify \cite{arechiga2014using}. We assume that the communication is not ideal and is subject to communication delays and package losses. Since the inter-vehicle communication is an ad-hoc wireless network where the packages are broadcasted to all neighboring vehicles, the package loss rate may increase in noisy environments, and a random delay may also increase in crowed environments \cite{he2013semi,dey2016review}. 

In summary, our main contribution in this paper lies on verifying a CACC control envelope to guarantee a passive safety property for non-ideal communication considering communication delays and package loss. Furthermore, we adopt  differential cut and weakening dL  [14], [15] axioms to reduce the verification complexity in KeYmaera [13]. To the best of our knowledge, it has not been attempted before.
\begin{itemize}
	\item Unlike \cite{ploeg2011design}, \cite{van2015safety} and \cite{santini2015consensus}, in which designed a regulator that is string stable, this work propose to verify a safety property for a control envelope. Thus, it ensures collision avoidance for a heterogeneous traffic scenario because any scenario that the regulators are in the control envelope will ensure the safety property, which it could be from semi-autonomous vehicles where the control systems intervene to avoid collisions to a fully autonomous vehicle which no human driving is required. 
	\item Unlike \cite{loos2011adaptive} and \cite{mitsch2012towards}, in which the verification considered an ideal network, this work considers a more realistic scenario where the communication is non-ideal and subject to delays and package loss. 
	\item Unlike \cite{loos2013efficiency}, in which the safety specification expected human intervention considering package loss, the control envelope considered in this paper does not require a human driver to ensure safety under non-deterministic package losses. Further, in this work, communication delays is also contemplated. 
	\item Unlike \cite{platzer2012structure} and \cite{platzer2015uniform}, in which the differential cut and weakening d$\mathcal{L}$ axioms were used to solve undecidable differential equations, in this work, those axioms were used to reduce the computational complexity of the verification in KeYmaera \cite{platzer2010logical}. Since d$\mathcal{L}$ verification is a semi-decidable decision problem \cite{platzer2010logical}, a procedure which reduces the axioms computation complexity can increase the class of verifiable systems in d$\mathcal{L}$.
\end{itemize}

The rest of the paper is organized as follows. Section \ref{sec:model} presents the scenario and the problem formulation of this work. Sections \ref{sec:formula}, \ref{sec:hybridprogram}, \ref{sec:loopinvariant}, \ref{sec:safe} and \ref{sec:verification} describe in details the verification approach. Finally, Section \ref{sec:conclusion} concludes the paper.

\section{System Model and Problem Formulation}\label{sec:model}

\begin{figure}
	\centering
	\includegraphics[width=0.7\linewidth]{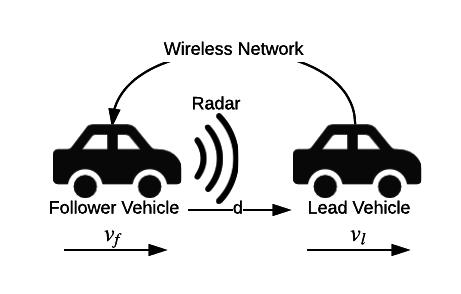}
	\caption{The local single lane highway scenario considered in this work.}
	\label{fig:highway-control-system---cacc2}
\end{figure}

This work considers a highway scenario illustrated in Fig.~\ref{fig:highway-control-system---cacc2}, in which it is assumed that all vehicles are equipped with inter-vehicle wireless network devices. The network can be ad-hoc, and all vehicles in the highway may broadcast network packages. Hence, it is assumed that the follower vehicles are equipped with a localization system that can distinguish the package from the vehicle immediately in front of it, i.e. the lead vehicle. Further, the follower vehicles are equipped with a radar sensor to read the relative distance $d$ to the lead vehicle. Both vehicles are equipped with sensors to read their current velocities (i.e. $v_f$ and $v_l$, follower and lead vehicle  velocities), and the inter-vehicle network allows the follower vehicle to receive the readings from lead vehicle velocity sensor. Finally, the network is not ideal and subject to communication delays and package losses. 

\begin{figure}
	\centering
	\includegraphics[width=0.7\linewidth]{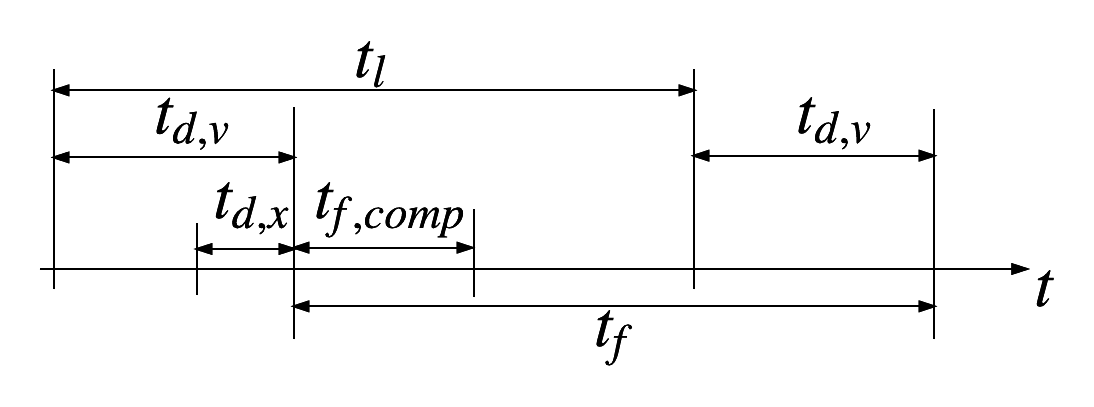}
	\caption{The delays considered in the proposed system.}
	\label{fig:highway-control-system---mac}
\end{figure}

The Fig.~\ref{fig:highway-control-system---mac} shows the delays in this scenario. Th delays are non-deterministic, and we know bounds. The lead vehicle will transmit its current velocity periodically with period $t_l$. We also consider that the control network may introduce a significant delay $t_{d,v}$, but its maximum value $\tau$ is known (i.e. $t_{d,v} \leq \tau$). The network receiving period $t_f$ is the time between lead velocity samples received by the follower vehicle, and its maximum value $\epsilon$ is known and always greater or equal to maximum communication delay (i.e. $\epsilon \geq \tau$). Thus, the follower vehicle can consider that the current network package is lost if $t_f > \epsilon$. The $t_{d,x}$ is a delay introduced by the relative distance sensor and is always smaller than the communication delay (i.e. $t_{d,x} \leq t_{d,v}$). The network period $t_l$ is greater enough to give time to the follower vehicle execute its control system $t_{f,comp}$ and smaller than $\epsilon - \tau$ (i.e. $t_{d,x} + t_{f,comp} \leq t_l \leq \epsilon - \tau$). Thus, the follower vehicle will receive new lead vehicle velocity value before or equal $\epsilon$ if the network package is not lost. 

The follower vehicle control system must guarantee that it will never actively collide with others vehicles, meaning that the follower vehicle will never be in an inevitable active collision state (IaCS) with another vehicle. An IaCS is an inevitable collision state (ICS) in which the agent is considered responsible for a collision. ICSs are system states which no matters the agent does in the future; it will not avoid the collision \cite{fraichard2004inevitable}. Hence, the ICS is relaxed with some assumptions which, in this case, are: all vehicles always drive in the same way (i.e. $v \geq 0$), change lane after the vehicle immediately behind in the other lane yields and are not responsible for a collision in the back. Hence, the follower vehicle can collide only if another vehicle enters in the lane without waiting for the follower vehicle to give way, is driving in wrong way ($v < 0$), or collides in the back of the follower vehicle. It is a fair safe requirement in a realistic traffic scenario because the traffic performance is increased assuming that all drivers respect some rules, including human and machine drivers. Therefore, the system must ensure a safety property that guarantees ICS-free longitudinal maneuvers between the follower and lead vehicle, where the lead vehicle is the vehicle immediately in front of the follower (i.e. always $d > 0$).

The continuous longitudinal dynamics is modeled by following differential equations: $x^{\prime}=v, v^{\prime}=a$, where $x$, $v$ and $a$ are the current vehicle position, velocity, and acceleration, respectively. These equations are considered adequate to describe longitudinal maneuvers  \cite{loos2011adaptive,mitsch2012towards,arechiga2012using,loos2013efficiency}. Hence, the follower vehicle dynamics is described by $x_f^{\prime}=v_f, v_f^{\prime}=a_f$ and the lead vehicle by $x_l^{\prime}=v_l, v_l^{\prime}=a_l$. The follower vehicle acceleration $a_f$ is changed after receiving a network package, and the lead vehicle changes its acceleration $a_l$ after sending a network package. We assume to know the maximum realizable braking deceleration $B$, and it is equal to both vehicles. We also assume to know the maximum acceleration $A$ and minimum braking power $b$ of the follower vehicle. 

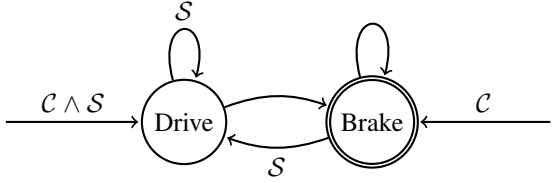
\begin{figure}[t]
	\centering
	\begin{tikzpicture}[shorten >=1pt,node distance=2.5cm,on grid,auto, bend angle=20, thick,scale=1, every node/.style={transform shape}]
	\node (s_0_1) {};
	\node[state] (s_2) [right=of s_0_1] {Drive};
	\node[state,accepting] (s_3) [right=of s_2] {Brake};
	\node (s_0_2) [right=of s_3] {};
	\path[->]
	(s_0_1) edge node [pos=0.5, sloped, above]{$\mathcal{C} \wedge \mathcal{S}$} (s_2)
	(s_0_2) edge node [pos=0.5, sloped, above]{$\mathcal{C}$} (s_3)
	(s_2) edge [loop above] node [pos=0.5, sloped, above]{$\mathcal{S}$} (s_2)
	(s_3) edge [loop above] node [pos=0.5, sloped, above]{} (s_3)
	(s_2) edge [bend left] node [pos=0.5, above]{} (s_3)
	(s_3) edge [bend left] node [pos=0.5, below]{$\mathcal{S}$} (s_2);
	\end{tikzpicture}
	\caption{Transition system that represents the control envelope to be found in the proposed problem.}
	\label{fig:controlenvelope}
\end{figure}

Instead of finding a unique control system, we propose to find a control envelope which ensures the safety property $d > 0$. A control envelope is a hybrid control system which represents a class of control systems \cite{arechiga2014using}. Fig.~\ref{fig:controlenvelope} shows this hybrid control system, where the initial constraint $\mathcal{C}$ is a first-order formula which constraints the states the system can start and the safe constraint $\mathcal{S}$ is a first-order formula which constraints the states the follower vehicle can drive. In the case that a vehicle is entering in front of the follower and at the same lane, a safe give way should respect the initial constraint $\mathcal{C}$. Being able to drive means that the vehicle can realize any acceleration $a$ in its domain $-B \leq a \leq A$. In case that this constraint is not true $\neg \mathcal{S}$, then the vehicle must brake, i.e. $-B \leq a \leq -b$. Now, we can formulate the problem.

\begin{problem}\label{problem1}
	Given the maximum network receiving period time $\epsilon > 0$, maximum communication delay $\tau \geq 0 \wedge \epsilon \geq \tau$, maximum braking deceleration $B > 0$ for both vehicles, minimum braking deceleration $b > 0 \wedge B \geq b$ and maximum acceleration $A > 0$ of the follower vehicle; find a control envelope that guarantees that the follower vehicle will never actively collide with a vehicle in the proposed highway scenario. 
\end{problem}

The approach used in this work to solve the Problem \ref{problem1} split the verification and design into five steps: design the d$\mathcal{L}$ formula $\phi_{cACC}$ that specifies the safety requirements, design the d$\mathcal{L}$ hybrid program $cACC$ that models this control envelope for the proposed cACC, find the loop invariant $\mathcal{C}_{loop}$ and the initial constraint $\mathcal{C}$ for the d$\mathcal{L}$ formula, find the safe constraint $\mathcal{S}$, and the verification of the resulting d$\mathcal{L}$ formula $\phi_{cACC}$. In the following sections, we will discuss each one of those steps.

\section{The d$\mathcal{L}$ formula $\phi_{cACC}$}\label{sec:formula}

The Differential Dynamic Logic $d\mathcal{L}$ verifies a symbolic hybrid system model, and, thus, can assist in verifying and finding symbolic parameters constraints. The iteration between the discrete and continuous dynamics is nontrivial and leads to nonlinear parameter constraints and nonlinearities in the dynamics. Most of the time, this turns into an undecidable problem for model checking \cite{platzer2010logical}. Hence, the model checking approach must rely on approximations. On the other hand, the $d\mathcal{L}$ uses a deductive verification approach to address infinite states; it does not rely on finite-state abstractions or approximations. Thus, a verification in $d\mathcal{L}$ can handle those nonlinear constraints. 

The $d\mathcal{L}$ formula is a specification language which is a first-order dynamic logic over the reals for hybrid systems. 

\begin{defn}[$d\mathcal{L}$ formulas]
	A $d\mathcal{L}$ formula \cite{platzer2010logical} ($\phi$ and $\psi$) is defined as:
	\begin{equation*}
	\phi, \psi ::= \chi \mid \neg \phi \mid \phi \wedge \psi \mid \forall x \phi \mid \exists x \phi \mid [\alpha] \phi \mid \langle \alpha \rangle \phi
	\end{equation*}
	where:
	\begin{itemize}
		\item $\alpha$ is a hybrid system modeled as a hybrid program which is defined in the Sec.~\ref{sec:hybridprogram}.
		\item $[\alpha] \phi$ holds true if $\phi$ is true after all runs of $\alpha$.
		\item $\langle \alpha \rangle \phi$ holds true if $\phi$ is true after at least one runs of $\alpha$.
	\end{itemize}
\end{defn}

The safety property proposed in this work is to ensure that the follower vehicle does not actively collide with another vehicle in a highway considering the scenario presented in Sec.~\ref{sec:model}. A collision will occur if the relative distance is zero (i.e. $d = 0$), and this distance can be related to the current position of the lead and follower vehicles (i.e. $d = x_l - x_f$). Since being a lead vehicle means that lead vehicle position is always greater and equal to follower position (i.e. $x_l \geq x_f$), the safety property which ensures that no collision occurs is $x_l - x_f > 0$. Further, we want ensures that the control envelope guarantees this property after any non-deterministic and a finite number of executions. Hence, the following d$\mathcal{L}$ formula $\phi_{cACC}$ specifies those requirements.

\begin{equation*}        
\phi_{cACC} \equiv \mathcal{C} \rightarrow [(cACC)^*] (x_l - x_f > 0)
\end{equation*}

\section{The d$\mathcal{L}$ hybrid program $cACC$}\label{sec:hybridprogram}
The hybrid systems are embedded to the d$\mathcal{L}$ as hybrid programs, a compositional program notation for hybrid systems. 

\begin{defn}[Hybrid Program]
	A hybrid program \cite{platzer2010logical} ($\alpha$ and $\beta$) is defined as:
	\begin{equation*}
	\alpha, \beta ::= \begin{cases}
	x_1 := \theta_1,...,x_n:=\theta_n \mid  ?\chi \mid \alpha ; \beta \mid \alpha \cup \beta \mid \alpha^* \mid \\
	x_1^{\prime} := \theta_1,...,x_n^{\prime}:=\theta_n \& \chi            
	\end{cases}  
	\end{equation*}
	where:
	\begin{itemize}
		\item $x$ is a state variable and $\theta$ a first-order logic term.
		\item $\chi$ is a first-order formula.
		\item $x_1 := \theta_1,...,x_n:=\theta_n$ are discrete jumps, i.e. instantaneous assignments of values to state variables. 
		\item $x_1^{\prime} := \theta_1,...,x_n^{\prime}:=\theta_n \& \chi$ is a differential equation system that represents the continuous variation in system dynamics. $x_i^{\prime} := \theta_i$ is the time derivative of state variable $x_i$, and $\& \chi$ is the evolution domain. 
		\item $?\chi$ tests a first-order logic at current state.
		\item $\alpha ; \beta$ is a sequential composition, i.e. the hybrid program $\beta$ will start after $\alpha$ finishes. 
		\item $\alpha \cup \beta$ is a nondeterministic choice.
		\item $\alpha^*$ is a nondeterministic repetition, which means that $\alpha$ will repeat for finite times. 
	\end{itemize}
\end{defn}

The proposed d$\mathcal{L}$ hybrid program $cACC$ that models the control envelope shown in Fig.\ref{fig:controlenvelope} is presented in Model~\ref{model:cACC}. It is assumed that the control system has a clock $t_f$ to detect a network package loss. This clock counts the maximum time between the time that the lead velocity was sampled, and it was received by the follower vehicle. Since the maximum communication delay $\tau$ is known, we know that this clock is equal to it at receiving event (i.e. $t_f = \tau$). The lead vehicle velocity received by the follower is the variable  $v_{l,d}$. Thus, the follower vehicle cannot directly access $v_l$ to take its decisions but only on $v_{l,d}$, and the safe constraints $safe_{delay}$ and $safe_{drop}$ can only depend on $v_{l,d}$ instead of $v_l$. 
\begin{model}\label{model:cACC}
	\caption{$cACC$ Hybrid Program}
	\begin{align*}
	cACC \equiv & (ctrl_f;dyn_{t};ctrl_l;dyn_{t_d}) \\
	ctrl_f \equiv & ctrl_{d} \cup ctrl_{pl} \cup ctrl_{s} \cup ctrl_{b} \\
	ctrl_{d} \equiv & (?((pkgdrop=0)\wedge(safe_{delay})); \\
	& t_f:=\tau;a_f:=*; ?(-B \leq a_f \leq A)) \\
	ctrl_{pl} \equiv & (?((pkgdrop=1) \wedge (safe_{drop})); \\
	& a_f:=*; ?(-B \leq a_f \leq A)) \\
	ctrl_{s} \equiv & (?(v_f=0);a_f:=0) \\
	ctrl_{b} \equiv & (a_f:=*; ?(-B \leq a_f \leq -b)) \\
	dyn_t \equiv & t:=0;\{x_f^{\prime} = v_f, v_f^{\prime} = a_f, x_l^{\prime} = v_l, v_l^{\prime} = a_l, \\
	& t^{\prime} = 1, t_f^{\prime} = 1 \& (v_f \geq 0 \wedge v_l \geq 0 \wedge t \leq \epsilon)\} \\
	ctrl_l \equiv & (pkgdrop:=0;v_{l,d}:=v_l)\cup (pkgdrop:=1); \\ 
	& a_l:=*; ?(a_l \geq -B) \\
	dyn_{t_d} \equiv & t_d:=0;\{x_f^{\prime} = v_f, v_f^{\prime} = a_f, \\
	& x_l^{\prime} = v_l, v_l^{\prime} = a_l, t_d^{\prime} = 1, t_f^{\prime} = 1 \\ 
	& \&  (v_l \geq 0 \wedge v_l \geq 0 \wedge t_d \leq \tau \wedge t_d+t\leq \epsilon)\}
	\end{align*}
\end{model}            

First, in the hybrid program $cACC$, the follower vehicle changes its acceleration $a_f$, i.e. executes $ctrl_f$. If it received a network package (i.e. $pkgdrop=0$) and it is safe to drive considering a network delay (i.e. $safe_{delay}$ holds true), or it detected a package loss (i.e. $pkgdrop=1$) and is safe even under this circumstance (i.e. $safe_{drop}$ holds true); then it can assign any acceleration in its domain (i.e. $ctrl_{d}$ or $ctrl_{pl}$). However, if it receive a network package, it will reset the $t_f$ clock (i.e. $t_f := \tau$). At any moment, the follower vehicle can brake (i.e. $ctrl_{b}$). And it is considered a mode to verify in the case that the follower vehicle maintain stopped (i.e. $ctrl_{s}$). Next, the dynamics $dyn_{t}$ models the continuous dynamics until the lead vehicle changes its acceleration $ctrl_l$. The lead vehicle control system is modeled by a control envelope that assumes that it can assign any acceleration that is greater than the maximum braking deceleration (i.e. $a_l \geq -B$) and send its current velocity (i.e. $v_{l,d}:=v_l$). However, we assume a network that may be non-deterministically subject to package loss (i.e. $(pkgdrop:=0;v_{l,d}:=v_l)\cup (pkgdrop:=1)$). Finally, the dynamics $dyn_{t_d}$ models the continuous dynamics after lead vehicle control $cltr_l$ execution and before the network package receiving or package loss events. 

However, the change in the dynamics of the lead vehicle during the follower cycle time increases the verification complexity for the KeYmaera. Normally, a verification in KeYmaera requires solving a real quantifier elimination (QE) at the end of the proof. Nevertheless, the QE complexity increases fast on the number of variables \cite{england2016complexity}. Since the lead vehicle control $cltr_l$ is executed asynchronously with follower control $ctrl_f$, the number of variables and equations of the final real QE increases and may not be computable. Therefore, we propose to augment the Model~\ref{model:cACC} using Differential Ghosts and Invariants \cite{platzer2012structure} to simplify the final QE problem. This approach is different from the original purpose which was to verify systems with undecidable differential equations without solving them; however, it offers an efficient procedure to simplify d$\mathcal{L}$ formulas that lead to complex QE problems.

The main idea is that QE become simpler for a double integrator system (i.e. $x^{\prime}=v,v^{\prime}=a$) when considering an acceleration that is an equality (e.g. $a=A$) instead of an inequality (e.g. $ -B \leq a \leq A$). Thus, it was proposed the following differential ghosts: ${x_{g,f}^{\prime}=v_{g,f},v_{g,f}^{\prime}=a_{g,f}}\&v_{g,f}\geq 0$ such that $x_{g,f} \geq x_f \wedge v_{g,f} \geq v_f$, and ${x_{g,l}^{\prime}=v_{g,l},v_{g,l}^{\prime}=a_{g,l}}\&v_{g,l}\geq 0$ such that $x_{g,l} \leq x_l \wedge v_{g,l} \leq v_l$. Since $x_{g,f} \geq x_f \wedge x_{g,l} \leq x_l \rightarrow x_l - x_f \geq x_{g,l} - x_{g,f}$, $x_{g,l} - x_{g,f} > 0 \rightarrow x_l - x_f > 0$. Therefore, introducing these ghosts will turn the proof simpler for KeYmaera because proving $x_{g,l} - x_{g,f} > 0$ leads to a simpler QE problem. 

Each control mode in $ctrl_f$ should assign an value for $a_{g,f}$ and $a_{g,l}$ to ensure that $x_{g,f} \geq x_f \wedge v_{g,f} \geq v_f$ and $x_{g,l} \leq x_l \wedge v_{g,l} \leq v_l$. Hence, the $ctrl_d$ and $ctrl_{pl}$ should assign $a_{g,f}=A$ and $a_{g,l}=-B$, the $ctrl_s$  $a_{g,f}=0$ and $a_{g,l}=-B$ and $ctrl_b$ $a_{g,f}=-b$ and $a_{g,l}=-B$. The resulting augmented d$\mathcal{L}$ hybrid program $cACC_{aug}$ is presented in Model~\ref{model:cACCaug}. The augmented dynamic models $dyn_{t}$ and $dyn_{t_d}$ considers when $v_{g,l} = 0$ letting it execute with $a_{g,l}=0$. Further, the case when $v_{g,f} = 0$ is considered because $v_{g,f} \geq v_f \wedge v_f \geq 0$. Hence, all reachable states in the hybrid program $cACC$ are in $cACC_{aug}$ (i.e. $\llbracket cACC \rrbracket \in \llbracket cACC_{aug} \rrbracket$). Therefore, if we prove $\phi_{cACC,aug} \equiv \mathcal{C} \rightarrow [cACC_{aug}] (x_l - x_f > 0)$, then we prove $\phi_{cACC}$ (i.e. $\phi_{cACC,aug} \rightarrow \phi_{cACC}$).
\begin{model*}\label{model:cACCaug}. 
	\caption{$cACC_{aug}$ Hybrid Program}
	\begin{align*}
	cACC_{aug} \equiv & (ghost_{init};ctrl_f;dyn_t;ctrl_l;dyn_{t_d}) \\
	ghost_{init} \equiv & x_{g,f}:=x_f;v_{g,f}:=v_f;x_{g,l}:=x_l;v_{g,l}:=v_l \\
	ctrl_f \equiv & ctrl_{d} \cup ctrl_{pl} \cup ctrl_{s} \cup ctrl_{b} \\
	ctrl_{d} \equiv & (?(pkgdrop=0);t_f:=\tau;?(safe_{delay}); a_f:=*; ?(-B \leq a_f \leq A);a_{g,f}:=A;a_{g,l}:=-B) \\
	ctrl_{pl} \equiv & (?(pkgdrop=1);?safe_{drop};a_f:=*; ?(-B \leq a_f \leq A);a_{g,f}:=A;a_{g,l}:=-B) \\
	ctrl_{s} \equiv & (?(v_f=0);a_f:=0;a_{g,f}:=0;a_{g,l}:=-B) \\
	ctrl_{b} \equiv & (a_f:=*; ?(-B \leq a_f \leq -b);a_{g,f}:=-b;a_{g,l}:=-B) \\
	dyn_{t} \equiv & t:=0;dyn_{t,ode};(?(v_{g,l}=0;a_{g,l}:=0;dyn_{t,ode})\cup(?(v_{g,l}>0)) \\    
	dyn_{t,ode} \equiv & \{x_f^{\prime} = v_f, v_f^{\prime} = a_f,x_l^{\prime} = v_l, v_l^{\prime} = a_l,x_{g,f}^{\prime} = v_{g,f}, v_{g,f}^{\prime} = a_{g,f},x_{g,l}^{\prime} = v_{g,l}, v_{g,l}^{\prime} = a_{g,l}, t^{\prime} = 1, t_f^{\prime}=1 \\
	& \& (v_f \geq 0 \wedge v_l \geq 0 \wedge v_{g,f} \geq 0 \wedge v_{g,l} \geq 0 \wedge t \leq \epsilon)\} \\
	ctrl_l \equiv & (pkgdrop:=0;v_{l,d}:=v_l)\cup (pkgdrop:=1);a_l:=*; ?(a_l \geq -B) \\
	dyn_{t_d} \equiv & t_d:=0;dyn_{t_d,ode};(?(v_{g,l}=0;a_{g,l}:=0;dyn_{t_d,ode})\cup(?(v_{g,l}>0)) \\    
	dyn_{t_d,ode} \equiv & \{x_f^{\prime} = v_f, v_f^{\prime} = a_f,x_l^{\prime} = v_l, v_l^{\prime} = a_l,x_{g,f}^{\prime} = v_{g,f},v_{g,f}^{\prime} = a_{g,f},x_{g,l}^{\prime} = v_{g,l}, v_{g,l}^{\prime} = a_{g,l}, t_d^{\prime} = 1, t_f^{\prime}=1 \\
	& \& (v_f \geq 0 \wedge v_l \geq 0 \wedge v_{g,f} \geq 0 \wedge v_{g,l} \geq 0 \wedge t_d \leq \tau \wedge t_d+t\leq \epsilon)\}
	\end{align*}
\end{model*}    

\section{The Loop Invariant $\mathcal{C}_{loop}$ and Initial Conditions $\mathcal{C}$}\label{sec:loopinvariant}    
The initial constraint $\mathcal{C}$ can be divided into two constraints: parameter constraint $\mathcal{C}_{par}$, which constraints the constants values, and initial state constraint $\mathcal{C}_{ini}$, which assign initial constraints to variables. The parameter constraint is specified by the assumptions defined in Sec.\ref{sec:model} and is $\mathcal{C}_{par} \equiv A > 0 \wedge B > 0 \wedge b > 0 \wedge B \geq b \wedge \epsilon > 0 \wedge \tau \geq 0 \wedge \epsilon \geq \tau$. The initial state constraint $\mathcal{C}_{ini}$ depends on the loop invariant $\mathcal{C}_{loop}$. 

The loop invariant $\mathcal{C}_{loop}$ is a first-order formula which permits to verify non-deterministic repetitions of a hybrid program (i.e. $\alpha^*$) \cite{platzer2010logical}. It should be initially valid (i.e. $\mathcal{C}_{par} \wedge \mathcal{C}_{ini} \rightarrow \mathcal{C}_{loop}$) and satisfy the safety property (i.e. $\mathcal{C}_{loop} \rightarrow x_l - x_f > 0$), and the hybrid program should preserve it (i.e. $\mathcal{C}_{par} \wedge \mathcal{C}_{loop} \rightarrow [cACC_{aug}] \mathcal{C}_{loop}$). To find it, we use an approach inspired from finding the controllability constraint presented in  \cite{platzer2010logical}. First, it was found the controllability constraint $\bar{\mathcal{C}}$ which the vehicle executing $ctrl_b$ is able to satisfy the safety property (i.e. $\bar{\mathcal{C}} \rightarrow [ghost_{init};ctrl_{b};dyn_t;ctrl_l;dyn_{t_d}](x_f - x_f > 0)$) because we assume in the control envelope that the follower vehicle must always be able to brake. Next, it was verified if this constraint is preserved when breaking (i.e. $\bar{\mathcal{C}} \rightarrow [ghost_{init};ctrl_{b};dyn_t;ctrl_l;dyn_{t_d}]\bar{\mathcal{C}}$). If so, the loop invariant should imply this constraint (i.e. $\mathcal{C}_{loop} \rightarrow \bar{\mathcal{C}}$). The controllability constraint $\bar{\mathcal{C}}$ found for the hybrid program $cACC_{aug}$ is $\bar{\mathcal{C}} \equiv a_l \geq -B \wedge v_f \geq 0 \wedge v_l \geq 0 \wedge x_l - x_f > 0 \wedge x_l - x_f > \frac{v_f^2}{2b} - \frac{v_l^2}{2B}$. It adds another constraint to prove which is simplified by using the proposed differential ghosts, i.e. $x_{g,l} - x_{g,f} > \frac{v_{g,f}^2}{2b} - \frac{v_{g,l}^2}{2B} \rightarrow x_l - x_f > \frac{v_f^2}{2b} - \frac{v_l^2}{2B}$  because $v_{g,f} \geq v_f \wedge v_{g,l} \leq v_l \wedge v_{g,l} \geq 0 \wedge v_f \geq 0  \rightarrow \frac{v_f^2}{2b} - \frac{v_l^2}{2B} \leq \frac{v_{g,f}^2}{2b} - \frac{v_{g,l}^2}{2B}$. 

The initial state constraint $\mathcal{C}_{ini}$ must take in account that $v_l$ is not direct accessible by the follower vehicle system; thus, it cannot be equal to controllability constraint $\mathcal{C}_{ini} \neq \bar{\mathcal{C}}$. However, $v_{l,d}$ depends on $v_l$ because we know the maximum delay $\tau$ and maximum deceleration $B$, and a constraint can be found to imply the controllability constraint (i.e. $\mathcal{C}_{ini} \rightarrow \bar{\mathcal{C}}$). Hence, since $v_{l,d} \geq 0$ and $v_{l,d} - B \cdot \tau \leq v_l$, $\big(v_{l,d} \geq B \cdot \tau \wedge x_l - x_f > \frac{v_f^2}{2b} - \frac{(v_{l,d} - B \cdot \tau)^2}{2B}\big) \vee  \big(v_{l,d} < B \cdot \tau \wedge x_l - x_f > \frac{v_f^2}{2b}\big) \rightarrow x_l - x_f > \frac{v_f^2}{2b} - \frac{v_l^2}{2B}$. If the package drop initially, it assumed that the current $v_{l,d}$ is zero. Thus, 
\begin{align*}
\mathcal{C}_{ini} \equiv & a_l \geq -B \wedge v_f \geq 0 \wedge v_l \geq 0 \wedge t_f \geq \tau \wedge x_l - x_f > 0 \wedge \\
& \Big[\Big(pkgdrop = 0 \wedge v_{l,d} - B \cdot \tau \leq v_l  \wedge v_{l,d} \geq 0 \\
& \wedge \big((v_{l,d} \geq B \cdot \tau \wedge x_l - x_f > \frac{v_f^2}{2b} - \frac{(v_{l,d} - B \cdot \tau)^2}{2B})  \\
& \vee (v_{l,d} < B \cdot \tau \wedge x_l - x_f > \frac{v_f^2}{2b})\big)\Big) \vee \\
& \Big(pkgdrop = 1 \wedge v_{l,d} = 0 \wedge x_l - x_f > \frac{v_f^2}{2b}\Big)\Big]
\end{align*}

Adding the premises about $v_{l,d}$, the loop invariant that implies the controllability constraint (i.e. $\mathcal{C}_{loop} \rightarrow \bar{\mathcal{C}}$) is,
\begin{align*}
\mathcal{C}_{loop} \equiv & a_l \geq -B \wedge v_f \geq 0 \wedge v_l \geq 0 \wedge t_f \geq \tau \wedge\\
& v_{l,d} \geq 0 \wedge x_l - x_f > 0 \wedge x_l - x_f > \frac{v_f^2}{2b} - \frac{v_l^2}{2B} \wedge \\
& \big((pkgdrop = 0 \wedge v_{l,d} - B \cdot \tau \leq v_l) \vee \\
& \phantom{\Big(}(pkgdrop = 1 \wedge v_{l,d} - B \cdot t_f \leq v_l)\big)
\end{align*}

\section{The Safe Condition $\mathcal{S}$}\label{sec:safe}
In the hybrid programs $cACC$ and $cACC_{aug}$ the drive mode presented in the control envelope (see Fig.~\ref{fig:controlenvelope}) was divided into two modes: drive under communication delay $ctrl_d$ and drive under package loss $ctrl_{pl}$. Hence, the resulting safe constraint is $\mathcal{S} \equiv safe_{delay} \vee safe_{drop}$, and it is possible to find them separately. Again, it was used an approach inspired by finding reactivity in \cite{platzer2010logical}. Therefore, we found the reactivity constraint for driving under communication delay $safe_{delay}$ such that $\mathcal{C}_{par} \wedge \mathcal{C}_{loop} \rightarrow [ghost_{init};ctrl_{d};dyn_t;ctrl_l;dyn_{t_d}] \mathcal{C}_{loop}$ is valid and for driving under package loss $safe_{drop}$ such that $\mathcal{C}_{par} \wedge \mathcal{C}_{loop} \rightarrow [ghost_{init};ctrl_{pl};dyn_t;ctrl_l;dyn_{t_d}] \mathcal{C}_{loop}$ is valid. It was found the same reactivity constraint found in previous works \cite{loos2011adaptive} (i.e. $x_l - x_f >  \frac{v_f^2}{2b} - \frac{v_l^2}{2B} + \big(\frac{A}{b} + 1\big)\big(\frac{A}{2}\epsilon^2 + \epsilon v_f\big)$); however, it is assumed that the follower vehicle only access $v_l$ through the variable $v_{l,d}$, and the following constraints consider the assumptions about $v_{l,d}$.    
\begin{align*}
safe_{delay} \equiv & \Big(v_{l,d} \geq B \cdot \tau \wedge x_l - x_f >  \frac{v_f^2}{2b} - \frac{(v_{l,d} - B \cdot \tau)^2}{2B} \\
&  + \big(\frac{A}{b} + 1\big)\big(\frac{A}{2}\epsilon^2 + \epsilon v_f\big)\Big) \vee \\
& \Big(v_{l,d} < B \cdot \tau \wedge x_l - x_f >  \frac{v_f^2}{2b} \\
& + \big(\frac{A}{b} + 1\big)\big(\frac{A}{2}\epsilon^2 + \epsilon v_f\big)\Big) \\
safe_{drop} \equiv & \Big(v_{l,d} \geq B \cdot t_f \wedge x_l - x_f >  \frac{v_f^2}{2b} - \frac{(v_{l,d} - B \cdot t_f)^2}{2B} \\
& + \big(\frac{A}{b} + 1\big)\big(\frac{A}{2}\epsilon^2 + \epsilon v_f\big)\Big) \vee \\
& \Big(v_{l,d} < B \cdot t_f \wedge x_l - x_f >  \frac{v_f^2}{2b} \\
& + \big(\frac{A}{b} + 1\big)\big(\frac{A}{2}\epsilon^2 + \epsilon v_f\big)\Big)
\end{align*}

\section{Verification of $\phi_{cACC}$}\label{sec:verification}
The approach applied to verify the formula $\phi_{cACC, aug}$ uses differential invariants and weakening axioms \cite{platzer2012structure,platzer2015uniform}. A differential invariant $\mathcal{D}$ is a first-order formula which represents properties of differential equations using an inductive technique. The differential invariant axiom states that $\mathcal{D}$ is a differential invariant with respect to a differential equation $\nu^{\prime} = f(\nu)$, where $\nu$ is the set of state variables in the system, if it is initially true (i.e. $\mathcal{P} \rightarrow \mathcal{D}$, where $\mathcal{P}$ is a set of premises), and its derivative $\mathcal{D}^{\prime}$ is valid (i.e. $\mathcal{P} \rightarrow \forall \nu ( \mathcal{D}^{\prime}|_{\nu^{\prime} = f(\nu)})$). The differential weakening is a d$\mathcal{L}$ axiom which internalizes that differential equations cannot leave their domain and differential invariants. This approach was originally used to solve differential equations which are not solvable \cite{platzer2008differential}. However, even though the differential equations in $cACC_{aug}$ are solvable in the differential dynamic logic, these solutions have equations that are not used in the proof and lack equations directly related to the properties that are needed to be proven. Therefore, the proposed approach solves as much as possible these properties using the differential invariants. 

To prove the formula $\phi_{cACC,aug}$, the differential equations in the hybrid program $cACC_{aug}$ must prove that it preserves the loop invariant (i.e. $\phi_{loop} \equiv \mathcal{C}_{par} \wedge \mathcal{C}_{loop} \rightarrow [cACC_{aug}] \mathcal{C}_{loop}$). To reduce the number of premises at the final QE problem, the goal is that the last the differential weakening must be able to generate all premises necessary to prove $\phi_{loop}$. 

Hence, for each differential equation system in the hybrid program $cACC_{aug}$, we apply differential cut and differential invariant axioms until it proves that the hybrid program preserves the loop invariant $\mathcal{C}_{loop}$ after applying differential weakening axiom. Differential equation systems in this hybrid program form three logical sequences: $\rho_1 = \{dyn_{t,ode} \xrightarrow{v_{g,l}>0} dyn_{t_d,ode} \xrightarrow{v_{g,l}>0} *\}$, $\rho_2 = \{dyn_{t,ode} \xrightarrow{v_{g,l}>0} dyn_{t_d,ode} \xrightarrow{v_{g,l}=0} dyn_{t_d,ode}\}$ and $\rho_3 = \{dyn_{t,ode} \xrightarrow{v_{g,l}=0} dyn_{t,ode} \rightarrow dyn_{t_d,ode} \xrightarrow{v_{g,l}=0} dyn_{t_d,ode}\}$. Thus, four differential invariants are designed: $\mathcal{D}_{t,1}$ for $dyn_{t,ode}$, $\mathcal{D}_{t,2}$ for $dyn_{t,ode}$ after $v_{l,g} = 0$ in $\rho_3$, $\mathcal{D}_{t_d,1}$ for $dyn_{t_d,ode}$ after $v_{g,l} > 0$ in $\rho_1$ and $\rho_2$, and $\mathcal{D}_{t_d,2}$ for $dyn_{t_d,ode}$ after $v_{l,g} = 0$ in $\rho_2$ and $\rho_3$. 

First, each constraint in $\mathcal{C}_{loop}$ is analyzed to find the differential invariants. The constraint $a_l \geq -B$ can be proven without a reasoning on the differential equations. The constraint $v_f \geq 0 \wedge v_l \geq 0$ can be proven using the differential equations domain. The constraint $t_f \geq \tau$ and $v_{l,d} \geq 0$ are differential invariants in all differential equations because is initially true and its differential is valid (i.e. $(t_f \geq \tau)^{\prime} \equiv 1 \geq 0$ and $(v_{l,d} \geq 0)^{\prime} \equiv 0 \geq 0$ are valid). The constraint $\mathcal{B}_{v_{l,d}} \equiv \big((pkgdrop = 0 \wedge v_{l,d} - B \cdot \tau \leq v_l) \vee (pkgdrop = 1 \wedge v_{l,d} - B \cdot t_f \leq v_l)\big)$ is not a differential invariant because the derivative is not valid (i.e. $(v_{l,d} - B \cdot \tau \leq v_l)^{\prime} \equiv (0 \geq a_l)$). However, the constraint $\bar{\mathcal{B}}_{v_{l,d}} \equiv \big((pkgdrop = 0 \wedge v_{l,d} - B \cdot t_d \leq v_l) \vee (pkgdrop = 1 \wedge v_{l,d} - B \cdot t_f \leq v_l)\big)$ is invariant for $dyn_{t_d,ode}$ if $v_{l,d} - B \cdot t_f \leq v_l$ is initially true. In fact, this constraint is invariant for $dyn_{t,ode}$ and, consequently, initially true for $dyn_{t_d,ode}$. Moreover, the constraint $\bar{\mathcal{B}}_{v_{l,d}}$ implies $\mathcal{B}_{v_{l,d}}$ (i.e. $\bar{\mathcal{B}}_{v_{l,d}} \rightarrow \mathcal{B}_{v_{l,d}}$) because $t_d \leq \tau$. The constraint $x_l - x_f > 0 \wedge x_l - x_f > \frac{v_f^2}{2b} - \frac{v_l^2}{2B}$ is not invariant for any differential equation. Nonetheless, it can be proven by using the proposed differential ghosts as discussed in Sec.~\ref{sec:hybridprogram} and \ref{sec:loopinvariant}. Thus, the differential invariants that helps to prove this constraint is the constraints $\mathcal{B}_{g} \equiv v_f \leq v_{g,f} \wedge v_l \geq v_{g,l} \wedge x_f \leq x_{g,f} \wedge x_l \geq x_{g,l}$ and $\mathcal{B}_{g}^{(0)} \equiv v_f \leq v_{g,f} \wedge x_f \leq x_{g,f} \wedge x_l \geq x_{g,l}$ after $v_{g,l} = 0$ because $v_l \geq v_{g,l}$ is not invariant in this situation, but true because $v_l \geq 0$. Further, we will need the solution for the differential ghosts variables $\mathcal{B}_{g,f}(t) \equiv v_{g,f} = v_{f,0} + a_{g,f} \cdot t \wedge x_{g,f} = x_{f,0} + v_{f,0} \cdot t + a_{g,f} \cdot \frac{t^2}{2}$, $\mathcal{B}_{g,l}(t) \equiv v_{g,l} = v_{l,0} + a_{g,l} \cdot t \wedge x_{g,l} = x_{l,0} + v_{l,0} \cdot t + a_{g,l} \cdot \frac{t^2}{2}$ and $\mathcal{B}_{g,l}^{(0)} \equiv v_{g,l} = 0 \wedge x_{g,l} = x_{l,0} + \frac{v_{l,0}^2}{2B}$ after $v_{g,l} = 0$, where $v_{f,0}$, $v_{l,0}$, $x_{f,0}$ and $x_{l,0}$ are the respective state variables values at beginning of the hybrid program. Therefore, the resulting differential invariants are:
\begin{align*}
\mathcal{D}_{t,1} \equiv & t \geq 0 \wedge t_f \geq \tau \wedge v_{l,d} \geq 0 \wedge v_{l,d} - B \cdot t_f \leq v_l \wedge \\
& \mathcal{B}_{g} \wedge \mathcal{B}_{g,f}(t) \wedge \mathcal{B}_{g,l}(t) \\
\mathcal{D}_{t,2} \equiv & t \geq 0 \wedge t_f \geq \tau \wedge v_{l,d} \geq 0 \wedge v_{l,d} - B \cdot t_f \leq v_l \wedge \\
& \mathcal{B}_{g}^{(0)} \wedge \mathcal{B}_{g,f}(t) \wedge \mathcal{B}_{g,l}^{(0)} 
\end{align*}
\begin{align*}
\mathcal{D}_{t_d,1} \equiv & t + t_d \geq 0 \wedge t_f \geq \tau \wedge v_{l,d} \geq 0 \wedge \bar{\mathcal{B}}_{v_{l,d}} \wedge \\
& \mathcal{B}_{g} \wedge \mathcal{B}_{g,f}(t+t_d) \wedge \mathcal{B}_{g,l}(t+t_d) \\
\mathcal{D}_{t_d,2} \equiv & t + t_d \geq 0 \wedge t_f \geq \tau \wedge \bar{\mathcal{B}}_{v_{l,d}} \wedge \\
& \mathcal{B}_{g}^{(0)} \wedge \mathcal{B}_{g,f}(t+t_d) \wedge \mathcal{B}_{g,l}^{(0)}.
\end{align*}

Since $a_{g,l} = -B$ in all control modes, the equations $\mathcal{B}_{g,f}(t+t_d)$ and $\mathcal{B}_{g,l}(t+t_d)$ imply $\mathcal{B}_{loop,1} \equiv x_{g,l} - x_{g,f} > 0 \leftrightarrow x_{l,0} - x_{f,0} > (a_{g,f} \cdot \frac{(t+t_d)^2}{2}+(t+t_d) \cdot v_{f,0}) - (v_{l,0} \cdot (t+t_d) - B \cdot \frac{(t+t_d)^2}{2})$ and $\mathcal{B}_{loop,2} \equiv x_{g,l} - x_{g,f} > \frac{v_{g,f}^2}{2b} - \frac{v_{g,l}^2}{2B} \leftrightarrow x_{l,0} - x_{f,0} >  \frac{v_{f,0}^2}{2b} - \frac{v_{l,0}^2}{2B} + (\frac{a_{g,f}}{b}+1)(a_{g,f} \cdot \frac{(t+t_d)^2}{2}+(t+t_d) \cdot v_{f,0})$. Further, the equations $\mathcal{B}_{g,f}(t+t_d)$ and $\mathcal{B}_{g,l}^{(0)}$ imply $\mathcal{B}_{loop,1}^{(0)} \equiv x_{g,l} - x_{g,f} > 0 \leftrightarrow x_{l,0} - x_{f,0} > (a_{g,f} \cdot \frac{(t+t_d)^2}{2}+(t+t_d) \cdot v_{f,0}) - \frac{v_{l,0}^2}{2B})$ and $\mathcal{B}_{loop,2}$. The resulting QE problems using the proposed differential ghost were found decidable in KeYmaera. On the other hand, if we do not use those ghosts, then we would have more complex constraints such as $v_l = v_{l,0} + a_{l,0} \cdot t + a_{l,1} \cdot t_d$ which depends on two different accelerations: the initial lead vehicle acceleration $a_{l,0} \geq -B$ and the acceleration after $ctrl_l$ $a_{l,1} \geq -B$. Those complex equations may turn the QE problem not computable. Therefore, the d$\mathcal{L}$ formula $\phi_{cACC,aug}$ is valid (and can be proved using KeYmaera \cite{platzer2010logical} with computable QE problems, and it proves $\phi_{cACC}$).

\begin{thm}\label{thm:safe}
	If the maximum network receiving period time $\epsilon > 0$, maximum communication delay $\tau \geq 0 \wedge \tau \leq \epsilon$, maximum braking deceleration $B > 0$ for both vehicles, minimum braking deceleration $b > 0 \wedge B \geq b$ and maximum acceleration $A > 0$ of the follower vehicle are known, then the control envelope shown in Fig.~\ref{fig:controlenvelope} with constraints $\mathcal{S} \equiv \big(pkgdrop  = 0 \wedge safe_d\big) \vee \big(pkgdrop = 1 \wedge safe_{pl}\big)$ and $\mathcal{C} \equiv \mathcal{C}_{par} \wedge \mathcal{C}_{ini}$ modeled by the hybrid program $cACC$ presented in Model~\ref{model:cACC} will never lead the follower vehicle to actively collide with another vehicle. The KeYmaera proof file may be downloaded from \url{ https://notredame.box.com/s/pc6zt6lx9tet16yyijdsaelzyqkpjt98}.
\end{thm}

\begin{example}
	A fully autonomous can be a regulator that accelerates until it reaches the maximum velocity of the road $V$ and, then, keeps this velocity. This regulator can be designed using any approach that assigns values to longitudinal acceleration $a$ that respects the constraints in the verified control envelope of the Theorem~\ref{thm:safe}. A simple example assigns values $a^*$ for the longitudinal acceleration such as,
	\begin{equation*}
		a^* = \begin{cases}
		A, & \text{if } \mathcal{S} \wedge v_f < V \\
		0, & \text{if } \mathcal{S} \wedge v_f = V \\
		-b, & \text{otherwise} \\
		\end{cases}
	\end{equation*}
\end{example}

\begin{example}
	Another example is control system that respects the constraints in the verified control envelope of the Theorem~\ref{thm:safe} is a semi-autonomous CACC system that assists a human drive such that the human is free to accelerate the vehicle when the constraint $\mathcal{S}$ holds true (i.e. $Manual$ driving). Otherwise, the controller take control of the vehicle and stops it (i.e. $Auto$ driving). This control system is illustrated as a transition system in Fig.~\ref{fig:semiautonomous}.
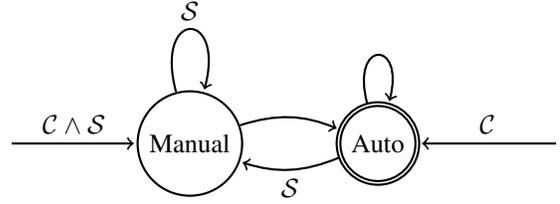
\begin{figure}[t]
	\centering
	\begin{tikzpicture}[shorten >=1pt,node distance=2.5cm,on grid,auto, bend angle=20, thick,scale=1, every node/.style={transform shape}]
	\node (s_0_1) {};
	\node[state] (s_2) [right=of s_0_1] {Manual};
	\node[state,accepting] (s_3) [right=of s_2] {Auto};
	\node (s_0_2) [right=of s_3] {};
	\path[->]
	(s_0_1) edge node [pos=0.5, sloped, above]{$\mathcal{C} \wedge \mathcal{S}$} (s_2)
	(s_0_2) edge node [pos=0.5, sloped, above]{$\mathcal{C}$} (s_3)
	(s_2) edge [loop above] node [pos=0.5, sloped, above]{$\mathcal{S}$} (s_2)
	(s_3) edge [loop above] node [pos=0.5, sloped, above]{} (s_3)
	(s_2) edge [bend left] node [pos=0.5, above]{} (s_3)
	(s_3) edge [bend left] node [pos=0.5, below]{$\mathcal{S}$} (s_2);
	\end{tikzpicture}
	\caption{Transition system that represents the control envelope to be found in the proposed problem.}
	\label{fig:semiautonomous}
\end{figure}
\end{example}

\section{Conclusion}\label{sec:conclusion}
In this paper, we proposed a safe control envelope for a realistic traffic scenario where communication delays and package loss may occur. Thus, any vehicle with a CACC system in the proposed control envelope is formally proven to avoid collisions with other vehicles which it can be blamed. It means that a collision can occur only in the back, or if the other enters the lane without waiting for a give way or drives in the wrong way. A fully self-driving vehicle as an autonomous vehicle which the human is allowed to drive and the machine intervenes only in unsafe states can be implemented using a CACC represented by this control envelope. Hence, the result presented here can take into account a heterogeneous traffic system. Moreover, the robustness to communication delays and package loss can address ad-hoc wireless network systems which each vehicle broadcast its packages. In such network systems, a random delay may be introduced to increase transmission success rate in a crowed environment. Thus, the impact of such delay to the safe distance can help to find tradeoffs between communication and control performances. Therefore, the CACC presented in this work is formally proven safe for a realistic traffic scenario.

In future work, we will implement a control system that uses this safe control envelope in simulated and real traffic scenarios. 

\small
\bibliographystyle{plain}
\bibliography{cACC_coDesign}

\end{document}